\documentstyle[preprint,eqsecnum,aps]{revtex}

\begin{document}
\draft

\title{Hamiltonian lattice quantum chromodynamics at finite density with Wilson fermions}

\author{Yi-Zhong Fang$^1$, 
and Xiang-Qian Luo$^{2,1}$\thanks{Corresponding author. Email address: stslxq@zsu.edu.cn}}
\address{
$^1$ Department of Physics, Zhongshan (Sun Yat-Sen) University, 
Guangzhou 510275, China \thanks{Mailing Address.} \\
$^2$ CCAST (World Laboratory), P.O. Box 8730, 
Beijing 100080, China
}

\date{\today}
\maketitle

\begin{abstract}
Quantum chromodynamics (QCD) at sufficiently high density is 
expected to undergo a chiral phase transition.
Understanding such a transition 
is of particular importance for neutron star or quark star physics.
In Lagrangian SU(3) lattice gauge theory, the standard approach breaks down 
at large chemical potential $\mu$,
due to the complex action problem. 
The Hamiltonian formulation of lattice QCD doesn't encounter such a problem.
In a previous work, we developed a Hamiltonian approach at
finite chemical potential $\mu$ and obtained reasonable results in the strong coupling regime.
In this paper, we extend the  previous work to Wilson fermions. 
We study the chiral behavior and calculate the vacuum energy, chiral condensate and quark
number density, as well as the masses of light hadrons.
There is a first order chiral phase transition at zero temperature. 
\end{abstract}



\section{Introduction}

Quantum Chromodynamics (QCD) is the fundamental theory of strong interactions.
It is a SU(3) gauge theory of quarks and gluons.
Precise determination of the QCD phase diagram on temperature $T$ and chemical
potential $\mu$ plane will provide valuable information for the experimental search for quark-gluon
plasma (QGP). The ultimate goal of machines like the Relativistic Heavy Ion Collider (RHIC)
at BNL and the Large Hadron Collider (LHC) at CERN is to create the QGP phase,
and replay the birth and evolution of the Universe. 
Such a new state of matter may also exist in the core of neutron stars or quark stars
at low temperature $T$ and large chemical potential $\mu$. Lattice gauge theory (LGT), 
proposed by Wilson\cite{Wilson:1974sk}
is a first principle non-perturbative method
for QCD. Although it is the most reliable technique for investigating phase transitions in QCD,
it is not free of problems:  complex action at finite chemical potential 
and species doubling with naive fermions.

In Lagrangian formulation of LGT at finite chemical potential, 
the success is limited to SU(2) gauge theory\cite{Hands:1999md,Aloisio:2000rb},
while in the physical SU(3) case, complex action\cite{Hasenfratz:1983ba,Palumbo:2002jf}
spoils numerical simulations with importance sampling.
Even though much effort\cite{Fodor:2002hs,deForcrand:2002ci,D'Elia:2002pj} 
has recently been made for SU(3) LGT,
and some very interesting information on the phase diagram at large $T$ and small $\mu$
has been obtained, 
it is still extremely difficult to do simulations at large chemical potential. 
QCD at large $\mu$ is of particular importance for neutron star or quark star physics.
Hamiltonian formulation of LGT doesn't encounter the notorious ``complex action problem''. 
Recently, we proposed a Hamiltonian approach 
to LGT with naive fermions
at finite chemical potential\cite{Gregory:1999pm,Luo:2000xi}
and solve it in the strong coupling regime. 
We predicted that at zero temperature, 
there is a first order chiral phase transition at critical chemical potential
$\mu_C=m_{dyn}^{(0)}=M_N^{(0)}/3$, with $m_{dyn}^{(0)}$ and $M_N^{(0)}$ being the 
dynamical mass of quark and nucleon mass at $\mu=0$ respectively.
(We expect this is also true for Kogut-Susskind fermions.) 
By solving the gap and Bethe-Salpeter equations, 
the authors of Ref. \cite{LeYaouanc:1987ff} obtained the critical point the same as ours;
but they concluded that the chiral transition is of second order, different from ours.
Our order of  transition is consistent with other lattice simulation results\cite{Aloisio:1999nk}.

Wilson's approach to lattice fermions\cite{Wilson:1974sk}
has been extensively used in hadron spectrum calculations 
as well as
in QCD at finite temperature.
It avoids the species doubling and preserves the flavor symmetry,
but it explicitly breaks the chiral symmetry\cite{Wilson:1974sk,Rothe,Montvay,Luo:fc,Chen:ej,Luo:1990dm,Azcoiti:1995bx}, 
one of the most important symmetries  of the original theory.
Non-perturbative fine-tuning of the bare fermion mass has to be done,
in order to define the chiral limit\cite{Smit:1980nf,Fang:2001ry}.

  In this paper, we study Hamiltonian lattice QCD with Wilson fermions at finite chemical potential. 
We derive the effective Hamiltonian in the strong coupling regime 
and diagonalize it by Bogoliubov transformation. 
The vacuum energy, chiral condensate, and masses of pseudo-scalar,
vector meson and nucleons are computed. 
In the non-perturbatively defined chiral limit,
we obtain reasonable results for the critical point and some physical quantities
in the large $N_c$ limit, 
with $N_c$ the number of colors.

To our knowledge,
the only existing literature about the same system ($r\not=0$ and $\mu\not=0$)
is  Ref.\cite{Umino:2001ie}, where the author
used a very different approach: the solution to the gap equation.
In contrary to the conventional predictions\cite{Smit:1980nf},
the author found that even at $\mu=0$, there is a critical value for the effective four fermion coupling $K$,
below which dynamical mass of quark vanishes. 
He introduced the concept of total chemical potential
and found the transition order depends on the input parameters $K$ and $r$ as well as the momentum.
In contrast, we find that in the chiral limit, 
dynamical mass of quark doesn't vanish for all values of $K$
if $\mu<\mu_C$ (the chiral-symmetry broken phase); and at  
$\mu=\mu_C$, our order of chiral phase transition doesn't depend on the input parameter.

  The rest of the paper is organized as follows. In Sec.\ref{our approach}, 
we derive the effective Hamiltonian at finite chemical potential. 
In  Sec.\ref{results}, we present the results for the vacuum energy, chiral condensate, 
and hadron masses. 
In Sec.\ref{critical},
we estimate the critical chemical potential at zero temperature. 
The results are summarized  in Sec.\ref{discussion}.

\section{Effective Hamiltonian in the strong coupling regime}
\label{our approach}

\subsection{The $\mu=0$ case}
\label{zero_potential}

We begin with QCD Hamiltonian\footnote{We use a representation of $\gamma$ matrices described 
in D. Lurie,¡¡
Particles and Fields, Interscience Publishers, John Wiley $\&$ Sons, Inc. (1968).}
with Wilson fermions at chemical potential $\mu=0$ 
on 1 dimensional continuum time and 3 dimensional spatial discretized lattice, 
\begin{eqnarray}
H &=& M \sum_{x} \bar{\psi}(x)\psi(x)
+{1 \over 2a}\sum_{x}\sum_{k=\pm1}^{\pm d}\bar{\psi}(x)\gamma_{k}U(x,k)\psi(x+{\hat k})
\nonumber \\
&-& {r \over 2a} \sum_{x}\sum_{k=\pm1}^{\pm d}\bar{\psi}(x)U(x,k)\psi(x+{\hat k})
+ {g^{2} \over 2a} 
\sum_{x}\sum_{j=1}^{d} E^{\alpha}_{j}(x)E^{\alpha}_{j}(x)\nonumber \\
&-& {1 \over ag^{2}} \sum_{p} {\rm Tr} \left(U_{p}+U_{p}^{+}-2\right),
\label{first}
\end{eqnarray}
where 
\begin{eqnarray}
M=m+ {rd \over a},
\end{eqnarray}
$d=3$ is the spatial dimension and $m$, $a$, $r$ and $g$ are respectively the bare fermion mass, 
spatial lattice spacing, 
Wilson parameter, and bare coupling constant. 
$U(x,k)$ is the gauge link variable at site $x$ and direction ${\hat k}$.
Fermion field $\psi$ on each lattice site carries spin (Dirac), color and flavor indexes; here and in the following, whenever there
is a summation sign ``$\sum_x$",  summations over these indexes are implied. 
The convention $\gamma_{-k}=-\gamma_{k}$ is used.  
$E_j^{\alpha}(x)$ is the color-electric field at site $x$ and direction $j$, and 
summation over $\alpha=1,2, ..., 8$ is implied.   
$U_p$ is the product of gauge link variables around an elementary spatial plaquette, and 
it represents the color magnetic interactions. 
In the continuum limit $a \to 0$, Eq.(\ref{first}) approaches 
to the continuum QCD Hamiltonian in the temporal gauge $A_{4}=0$.

The effective Hamiltonian, obtained by strong coupling expansion up to the second order,
is\cite{Fang:2001ry}
\begin{eqnarray}
H_{eff} &=& M\sum_{x}\bar{\psi}_f(x)\psi_f (x)-{K(r^{2}+1)d \over a}
\sum_{x}\psi_f^{\dagger}(x)\psi_f (x)
\nonumber \\
&+&\frac {K}{8aN_c}\sum_{x}
\sum_{k=\pm j}
\bigg( (r^{2}+1)
\psi_{f_1}^{\dagger}(x)\psi_{f_2} (x) \psi_{f_2}^{\dagger}(x+{\hat k})\psi_{f_1} (x+{\hat k}) 
\nonumber \\
&+& (r^{2}-1)\psi_{f_1}^{\dagger} (x)\gamma_{4}\psi_{f_2} (x)
\psi_{f_2}^{\dagger}(x+{\hat k})\gamma_{4}\psi_{f_1}(x+{\hat k})
\nonumber \\
&- & (r^{2}-1)\psi_{f_1}^{\dagger}(x)\gamma_{5}\psi_{f_2} (x)\psi_{f_2}^{\dagger}(x+{\hat k})\gamma_{5}
\psi_{f_1} (x+{\hat k})
\nonumber \\
&+&
(r^{2}+1)\psi_{f_1}^{\dagger}(x)\gamma_{4}\gamma_{5}\psi_{f_2} (x)
 \psi_{f_2}^{\dagger}(x+{\hat k})\gamma_{4}\gamma_{5}\psi_{f_1} (x+{\hat k})
\nonumber \\
& +& 
\left(r^{2}+(1-2\delta_{\vert k \vert, j})\right)
\psi_{f_1}^{\dagger}(x)\gamma_{4}\gamma_{j}\psi_{f_2} (x)
\psi_{f_2}^{\dagger}(x+{\hat k})\gamma_{4}\gamma_{j}\psi_{f_1} (x+{\hat k}) 
\nonumber \\
&- & 
\left( r^{2}-(1-2\delta_{\vert k \vert, j})\right)\psi_{f_1}^{\dagger}(x)\gamma_{j}\psi_{f_2} (x)
\psi_{f_2}^{\dagger}(x+{\hat k})\gamma_{j}\psi_{f_1} (x+{\hat k}) 
\nonumber \\
&- & 
\left(r^{2}+(1-2\delta_{\vert k \vert, j})\right) \psi_{f_1}^{\dagger}(x)\gamma_{4}\sigma_{j}\psi_{f_2} (x)
\psi_{f_2}^{\dagger}(x+{\hat k})\gamma_{4}\sigma_{j}\psi_{f_1} (x+{\hat k})
\nonumber \\
&-& \left(r^{2}-(1-2\delta_{\vert k \vert, j})
\right)\psi_{f_1}^{\dagger}(x)\sigma_{j}\psi_{f_2} (x)\psi_{f_2}^{\dagger}(x+{\hat k})
\sigma_{j}\psi_{f_1} (x+{\hat k})\bigg).
\label{third}
\end{eqnarray}
This effective Hamiltonian is equivalent to that in Ref. \cite{Smit:1980nf}, though different representations of $\gamma$ matrices are used.
Here $\sigma_{j}=\epsilon_{jj_{1}j_{2}}\gamma_{j_{1}}\gamma_{j_{2}}$. 
The flavor indexes are explicitly written. 
This effective Hamiltonian describes the nearest-neighbor four fermion interactions,
with 
\begin{eqnarray}
K=\frac {1}{g^{2}C_N}
\end{eqnarray}
being the effective four fermion coupling constant. Here $C_N=(N_c^2-1)/(2N_c)$ is the Casimir invariant of the
SU$(N_c)$  gauge group.

\subsection{The $\mu \ne 0$ case}

In the continuum, 
the grand canonical partition function of QCD at finite temperature $T$ and chemical potential $\mu$ 
is
\begin{eqnarray}
Z={\rm Tr} ~{\rm e}^{- \beta \left( H - \mu N \right)}, ~~~ 
\beta = (k_{B} T)^{-1} , 
\label{sixth}
\end{eqnarray}
where $k_B$ is the Boltzmann constant
and $N$ is particle number operator
\begin{eqnarray}
N=\sum\limits_{x} \psi^{\dagger}(x) \psi (x).
\label{seventh}
\end{eqnarray}
According to Eq.(\ref{sixth}) and following the procedure in Sec.\ref{zero_potential}, 
the role of the Hamiltonian at strong coupling is now played by
\begin{eqnarray}
H^{\mu}_{eff}=H_{eff}- \mu N .
\label{eighth}
\end{eqnarray}
In this Hamiltonian, there are three input parameters: $r$, $m$ and $\mu$. Suppose we study the phase structure
of the system in the chiral limit.  Such a limit can be reached by fine-tuning the bare quark mass $m$
so that the pion becomes massless.
In such a case, there are only two free parameters left: $r$ and $\mu$.

The vacuum energy is the expectation value of $H-\mu N$ in its ground state $\vert \Omega \rangle$, and 
also the expectation value of $H^{\mu}_{eff}$ in its ground state $\vert \Omega_{eff} \rangle$, 
given by 
\begin{eqnarray}
E_{\Omega} &=& \langle \Omega \vert H - \mu N \vert \Omega \rangle =
\langle \Omega_{eff} \vert H ^{\mu}_{eff} \vert \Omega_{eff} \rangle .
%
\label{vacuum_energy}
\end{eqnarray}
%

\section{Physical quantities at $\mu \ne 0$ and $T=0$}
\label{results}

\subsection{Meson masses}
 
One way to compute the masses of mesons 
as well as the contributions of the mesons to the vacuum energy,  
is to bosonize the effective Hamiltonian Eq. (\ref{third}).
We introduce the following operators \cite{Gregory:1999pm,Luo:kk,Guo:xa}
\begin{eqnarray}
\Pi_{f_{1}f_{2}}(x) &=& \frac {1}{2\sqrt{-{\bar v}}}
\psi^{\dagger}_{f_{1}}(x)
(1-\gamma_{4})\gamma_{5}\psi_{f_{2}}(x), 
\nonumber \\
\Pi^{\dagger}_{f{2}f{1}}(x) &=& \frac {1}{2\sqrt{-{\bar v}}}\psi^{\dagger}_{f_{2}}
(x)(1+\gamma_{4})\gamma_{5}\psi_{f_{1}}(x), 
\nonumber \\
V_{jf_{1}f_{2}}(x) &=& \frac {1}{2\sqrt{-{\bar v}}}\psi^{\dagger}_{f_{1}}(x)(1-
\gamma_{4})\gamma_{j}\psi_{f_{2}}(x), 
\nonumber \\
V^{\dagger}_{jf_{2}f_{1}} (x) &=& \frac {1}{2\sqrt{-{\bar v}}}\psi^{\dagger}_
{f_{2}}(x)(1+\gamma_{4})\gamma_{j}\psi_{f_{1}}(x).
\label{twenty-seventh}
\end{eqnarray}
$j$ stands for the positive spatial direction, and ${\bar v}$ and $v^{\dagger}$ denote respectively expectation value  
of ${\bar \psi} \psi$ and $\psi^{\dagger} \psi$ in the vacuum state $\vert \Omega_{eff} \rangle$ of $H_{eff}$, i.e.,
\begin{eqnarray}
{\bar v} &=& \bigg\langle  {\bar \psi} (x) \psi (x) \bigg\rangle_{eff}= {1 \over N_fN_s} 
\langle  \Omega_{eff} \vert \sum_x {\bar \psi}(x) \psi (x) \vert \Omega_{eff} \rangle,
\nonumber\\
v^{\dagger} &=& \bigg\langle  \psi^{\dagger}(x) \psi (x) \bigg\rangle_{eff}
=
{1 \over N_fN_s} 
\langle \Omega_{eff} \vert  \sum_x \psi^{\dagger}(x) \psi (x) \vert \Omega_{eff} \rangle. 
\label{vv}
\end{eqnarray}
Here $N_s$ is the total number of lattice sites and $N_f$ the number of flavors.
It is shown in Appendix \ref{APPEN_BOSON}
 that under the linearization prescription\cite{Guo:xa}, 
operators defined in Eq. (\ref{twenty-seventh}) satisfy the canonical commutation relations for bosons and
then the effective Hamiltonian $H^{\mu}_{eff}$ in Eq.(\ref{eighth}) can be expressed 
in terms of these operators in the following way
\begin{eqnarray}
H^{\mu}_{eff} &\sim &  H^{\mu}_{Linear} = E_{\Omega}^{(0)} +H_{\Pi}+H_{V},
\label{bosonized}
\end{eqnarray}
where
\begin{eqnarray}
E_{\Omega}^{(0)} 
&=&
N_f N_s \bigg[ M{\bar v}- \left( {Kd  \left(1+ r^2 \right) \over a} + \mu \right) v^{\dagger}
+ {Kd r^2  \over 4aN_c} \left(v_2^{\dagger}+{\bar v_2}\right)
\nonumber \\
&-& {Kd \over 4aN_c} \left({\bar v_2}-v_2^{\dagger}+v_{2\sigma_j}^{\dagger}+{\bar v}_{2\sigma_j}\right) 
\bigg],
\label{energy_leadingNc}
\end{eqnarray}
and
\begin{eqnarray}
H_{\Pi} &=& \left(2M - {Kd (1-r^2) \over aN_c} {\bar v} \right)
\sum_{x,f_{1},f_{2}}\Pi^{\dagger}_{f_{2}f_{1}}(x)\Pi_{f_{1}f_{2}}(x)
\nonumber\\ &+& { K r^2 \over 4aN_c}{\bar v} \sum_{x,f_{1},f_{2},k}
\left(\Pi^{\dagger}_{f_{1}f_{2}}(x)\Pi_{f_{2}f_{1}}(x+{\hat k})
+\Pi_{f_{1}f_{2}}(x)\Pi^{\dagger}_{f_{2}f_{1}}(x+{\hat k})\right) 
\nonumber \\
&-& 
{K \over 4aN_c}{\bar v}
\sum_{x,f_{1},f_{2},k}\left(\Pi^{\dagger}_{f_{1}f_{2}}(x)
\Pi^{\dagger}_{f_{2}f_{1}}(x+{\hat k})+\Pi_{f_{1}f_{2}}(x)\Pi_{f_{2}f_{1}}(x+{\hat k})
\right),
\nonumber \\
H_{V} &=& \left(2M- {Kd (1-r^2) \over aN_c} {\bar v}\right) \sum_{x,f_{1},f_{2},j}
V^{\dagger}_{jf_{2}f_{1}}(x)V_{jf_{1}f_{2}}(x) 
\nonumber \\
&+& 
{Kr^2 \over 4aN_c}{\bar v}
\sum_{x,f_{1},f_{2},k,j}
\left(V^{\dagger}_{jf_{1}f_{2}}(x)V_{jf_{2}f_{1}}(x+{\hat k})
+V_{jf_{1}f_{2}}(x)V^{\dagger}_{jf_{2}f_{1}}(x+{\hat k})\right) 
\nonumber \\
&-& 
{K \over 4aN_c} 
{\bar v} \sum_{x,f_{1},f_{2},k,j}\left(V^{\dagger}_{jf_{1}f_{2}}(x)
V^{\dagger}_{jf_{2}f_{1}}(x+{\hat k})+V_{jf_{1}f_{2}}(x)V_{jf_{2}f_{1}}(x+{\hat k})\right)
(1-2\delta_{\vert k \vert, j}).
\label{thirty-first}
\end{eqnarray}
$v_2^{\dagger}$, ${\bar v}_2$, $v_{2\sigma_j}^{\dagger}$ and ${\bar v}_{2\sigma_j}$ 
are expectation value of  four fermion operators
\begin{eqnarray}
{\bar v}_2 &=&
{1 \over 2dN_fN_s} \langle \Omega_{eff} \vert \sum_{x,f_1,f_2,k}{\bar \psi}_{f_1} (x) \psi_{f_2} (x) 
{\bar \psi}_{f_2}(x+{\hat k}) \psi_{f_1} (x+{\hat k}) \vert \Omega_{eff} \rangle ,
\nonumber \\
v_2^{\dagger} &=&
{1 \over 2dN_fN_s} \langle \Omega_{eff} \vert \sum_{x,f_1,f_2,k}\psi_{f_1}^{\dagger} (x) \psi_{f_2} (x) 
\psi_{f_2}^{\dagger}(x+{\hat k}) \psi_{f_1} (x+{\hat k}) \vert \Omega_{eff} \rangle ,
\nonumber \\
{\bar v}_{2\sigma_j} &=&
{1 \over 2dN_fN_s} \langle \Omega_{eff} \vert 
\sum_{x,f_1,f_2,k,j}\left(r^2+\left(1-2\delta_{\vert k \vert,j} \right)\right){\bar \psi}_{f_1} (x)\sigma_j \psi_{f_2} (x) 
{\bar \psi}_{f_2}(x+{\hat k}) \sigma_j \psi_{f_1} (x+{\hat k}) \vert \Omega_{eff} \rangle ,
\nonumber \\
v_{2\sigma_j}^{\dagger} &=&
{1 \over 2dN_fN_s} \langle \Omega_{eff} \vert 
\sum_{x,f_1,f_2,k,j}\left(r^2-\left(1-2\delta_{\vert k \vert,j} \right)\right)\psi_{f_1}^{\dagger} (x) \sigma_j \psi_{f_2} (x) 
\psi_{f_2}^{\dagger}(x+{\hat k}) \sigma_j \psi_{f_1} (x+{\hat k}) \vert \Omega_{eff} \rangle .
\nonumber \\
\label{v2}
\end{eqnarray}

After a Fourier transformation
\begin{eqnarray}
\Pi_{f_{1}f_{2}}(x)=\sum\limits_{p}e^{ipx} {\tilde \Pi}_{f_{1}f_{2}}(p),
\label{thirty-second}
\end{eqnarray}
$H_{\Pi}$ in Eq.(\ref{thirty-first}) becomes
\begin{eqnarray}
H_{\Pi} &=& \left(2M-{Kd \over aN_c}(1-r^{2}) \right)
\sum_{p,f_{1},f_{2}} {\tilde \Pi}^{\dagger}_{f_{1}f_{2}} (p) {\tilde \Pi}_{f_{2}f_{1}}(p) 
\nonumber \\
&+& {Kr^{2} \over 2aN_c}{\bar v} \sum_{f_{1}f_{2}}\sum_{p} 
\left(
{\tilde \Pi}^{\dagger}_{f_{1}f_{2}} (p) {\tilde \Pi}_{f_{2}f_{1}}(p)
+ {\tilde \Pi} (p)_{f_{1}f_{2}} {\tilde \Pi}^{\dagger}_{f_{2}f_{1}}(p) \right)
\sum_{j=1}^d \cos{p_{j}a} 
\nonumber \\
&-& 
{K \over 2aN_c} {\bar v} \sum_{p,f_{1},f_{2}} 
\left(
{\tilde \Pi}^{\dagger}_{f_{1}f_{2}} (p) {\tilde \Pi}^{\dagger}_{f_{2}f_{1}} (-p)
+ {\tilde \Pi}_{f_{1}f_{2}} (-p) {\tilde \Pi}_{f_{2}f_{1}}(p)
\right)
\sum_{j=1}^d \cos{p_{j}a}.
\label{twentyforth}
\end{eqnarray}
The Bogoliubov transformation\cite{Fang:2001ry}
\begin{eqnarray}
{\tilde \Pi} (p) & \rightarrow & {\tilde \Pi} (p)\cosh{u_p}+{\tilde \Pi}^{\dagger} (-p)\sinh{u_p} , 
\nonumber \\
{\tilde \Pi}^{\dagger} (p) & \rightarrow & {\tilde \Pi}^{\dagger}(p)\cosh{u_p}+{\tilde \Pi}(-p)\sinh{u_p},
\label{twentyfifth}
\end{eqnarray}
diagonalizes  $H_{\Pi}$ if 
\begin{eqnarray}
\tanh{2u_{p}} &=& -\frac {2G_{2}}{G_{1}}\sum_{l=1}^d\cos{p_{l}a} ,
\nonumber\\
G_{1} &=& 2M-\frac {Kd}{aN_c}{\bar v}(1-r^{2})+\frac {Kr^{2}}{aN_c}{\bar v}
\sum_{l=1}^d\cos{p_{l}a},
\nonumber \\
G_{2} &=& -\frac {K}{2aN_c}{\bar v}.
\label{thirty-forth}
\end{eqnarray}
The resulting $H_{\Pi}$ is
\begin{eqnarray}
H_{\Pi} 
&=& 
G_{1}\sum_{p,f_{1},f_{2}}(1-\tanh^{2}{2u_{p}})^{\frac {1}{2}}
{\tilde \Pi}^{\dagger}_{f_{1}f_{2}}(p) {\tilde \Pi}_{f_{2}f_{1}} (p) 
\nonumber \\
&-& 
\frac {G_{1}}{2}N^{2}_{f}\sum_{p}\left(\left(1-(1-\tanh^{2}{2u_{p}})^{\frac {1}{2}}\right) 
+\frac {2G_{2}r^{2}}{G_{1}}\sum_{l=1}^d\cos{p_{l}a}\right).
\label{thirty-third}
\end{eqnarray}
Now ${\tilde \Pi}^{\dagger}_{f_{1}f_{2}}(p)$ stands for the pseudo-scalar
creation operator in the momentum space.
According to Eq. (\ref{thirty-third}), 
the difference between the pseudo-scalar meson energy and vacuum energy is
\begin{eqnarray}
E_{\Pi} &=& G_{1} \left(1-\tanh^{2}{2u_p}\right)^{\frac {1}{2}}
\nonumber \\
&=& \left(2M-\frac {Kd}{aN_c}{\bar v}(1-r^{2})-{K \over aN_c} (1-r^{2}){\bar v}\sum_{l=1}^d
\cos{p_{l}a}\right)^{\frac {1}{2}}
\nonumber \\
& \times & \left(2M-\frac {Kd}{aN_c}{\bar v}(1-r^{2})
+{K \over aN_c}(1+r^{2}){\bar v} \sum_{l=1}^d\cos{p_{l}a}\right)^{\frac {1}{2}},
\label{thirty-fifth}
\end{eqnarray}
which gives the pseudo-scalar mass when $p_j = 0$.
The pseudo-scalar mass square is
\begin{eqnarray}
M_{\Pi}^2=E_{\Pi}^2 \vert_{p_j=0} &=& 4 \left(M+\frac {Kdr^2}{aN_c}{\bar v}\right)
\left(M+\frac {Kdr^2}{aN_c}{\bar v}-\frac {Kd}{aN_c}{\bar v}\right).
\label{PCAC}
\end{eqnarray}
In order to define the chiral limit, one
has to fine tune $M \to M_{chiral}$ so that the pion becomes massless. 
From Eq. (\ref{PCAC}), we get
\begin{eqnarray}
M_{chiral}=-\frac {Kdr^{2}}{aN_c}{\bar v}.
\label{thirty-sixth}
\end{eqnarray}
In this limit,  the pseudo-scalar mass square behaves as $M_{\Pi}^2 \propto M-M_{chiral}$,
which is the PCAC relation.

The $H_{V}$ sector in Eq. (\ref{thirty-first}) can be considered in a similar way.  After a Fourier transformation
\begin{eqnarray}
V_{j}(x)=\sum_{p}e^{ipx}{\tilde V}_j (p)
\label{thirty-seventh}
\end{eqnarray}
and a Bogoliubov transformation
\begin{eqnarray}
{\tilde V}_j (p) & \rightarrow &  {\tilde V}_j (p)\cosh{w^{(j)}_p} +{\tilde V}^{\dagger}_j(-p)\sinh{w^{(j)}_p},
\nonumber \\
{\tilde V}^{\dagger}_j (p) & \rightarrow & {\tilde V}^{\dagger}_j (p) \cosh{w_j(p)} +
{\tilde V}_j (-p) \sinh{w^{(j)}_p},
\label{thirty-eighth}
\end{eqnarray}
$H_V$ becomes a diagonalized one
\begin{eqnarray}
H_{V} &=& 
G_{1}\sum_{p,j,f_1,f_2} \left(1-\tanh^2 2w^{(j)}_p\right)^{{1 \over 2}}{\tilde V}^{\dagger}_{jf_{1}f_{2}} (p) V_{jf_{2}f_{1}} (p)
\nonumber \\
&-& {G_1 \over 2} N^2_f \sum_{p,j}
\left(\left(1-\left(1-\tanh^{2} 2w^{(j)}_p\right)^\frac {1}{2}\right)
+\frac {2G_{2}}{G_{1}}r^{2}\sum_{l=1}^d\cos{p_{l}a}\right),
\label{G1G2}
\end{eqnarray}
if 
\begin{eqnarray}
\tanh{2w^{(j)}_p}=-\frac {2G_{2}}{G_{1}}(\sum_{l=1}^d\cos{p_{l}a}-
2\cos{p_j}a).
\label{thirty-ninth}
\end{eqnarray}
Now ${\tilde V}^{\dagger}_{jf_{1}f_{2}}(p)$ stands for the vector
creation operator in the momentum space.
The vector mass  is 
\begin{eqnarray}
M_{V} = G_{1} \left(1-\tanh^2 2w^{(j)}_0\right)^{{1 \over 2}} 
\to^{M\to M_{c}}
-{2K \sqrt{d-1} \over aN_c} {\bar v}.
\end{eqnarray}

According to Eqs. (\ref{bosonized}),  (\ref{thirty-third}) and (\ref{G1G2}), the  vacuum energy reads
\begin{eqnarray}
E_{\Omega} &=& \langle \Omega \vert H^{\mu} \vert \Omega \rangle
\nonumber\\
&=& E_{\Omega}^{(0)}
- \frac {G_{1}}{2}N^{2}_{f}\sum_{p}\left(\left(1-(1-\tanh^{2}{2u_p})
^{\frac {1}{2}}\right) + \frac {2G_{2}r^{2}}{G_{1}}\sum_{l=1}^d \cos{p_{l}a}\right)
\nonumber\\
&-& \frac {G_1}{2}N^2_f\sum_{p,j}\left(\left(1-(1-\tanh^{2}2w^{(j)}_p)^\frac {1}{2}\right)
+\frac {2G_{2}}{G_{1}}r^{2}\sum_{l=1}^d\cos{p_{l}a}\right).
\label{vacuum_energyN_c}
\end{eqnarray}
This also gives the
thermodynamic potential (grand potential) at $T=0$.

As shown in Ref. \cite{Fang:2001ry}, at $\mu=0$ the results above 
are consistent with those of Smit in Ref. \cite{Smit:1980nf} where the $1/N_c$ expansion was used.

\subsection{Results in the large $N_c$ limit}

As shown in Appendix \ref{APPEN_VACUUM}, in the chiral limit,
the dominant contributions to the vacuum energy for large $N_c$ is
\begin{eqnarray}
E_{\Omega} & \to & 
N_f N_s \bigg[ M_{chiral} {\bar v}- \left( {Kd  \left(1+ r^2 \right) \over a} + \mu \right) v^{\dagger}
+ {Kd r^2  \over 4aN_c} \left( \left(v^{\dagger}\right)^2 +{\bar v}^2 \right)
-{Kd \over 4aN_c} \left({\bar v}^2-  \left( v^{\dagger}\right)^2 \right) 
\bigg] .
\nonumber\\
\label{vacuum_energy_large_Nc}
\end{eqnarray}

As shown Appendix \ref{APPEN_MEAN}, under the mean-field approximation, i.e.,
by Wick-contracting a pair of fermion fields
in the four fermion terms in Eq. (\ref{third}),  one can obtain
a bilinear Hamiltonian in the large $N_c$ limit
\begin{eqnarray}
H_{eff} \sim H_{MFA}=A\sum_{x} \bar{\psi} (x) \psi (x) +B\sum_{x} \psi^{\dagger} (x) \psi (x) +C ,
\label{meanfield}
\end{eqnarray}
where
\begin{eqnarray}
A &=& M_{chiral}-\frac {Kd}{2aN_{c}}(1-r^2) {\bar v},
\nonumber\\
B &=& \frac {Kd (1+r^2)}{a} \left( { v^{\dagger} \over 2N_{c}} -1\right),
\nonumber\\
C &=& -\frac {Kd}{4aN_{c}} 
\left( (1+r^2) {v^{\dagger}}^2 
- (1-r^2){\bar v}^2 \right) N_sN_f.
\label{consts}
\end{eqnarray}
The coefficient $A$ plays the role of dynamical mass of quark. 
According to Eq. (\ref{vacuum_energy}) and  Eq. (\ref{meanfield}), 
the vacuum energy in presence of the chemical potential $\mu$ is now
$E_{\Omega}=\langle \Omega_{eff} \vert H _{MFA} -\mu N \vert \Omega_{eff} \rangle$, which
agrees with Eq. (\ref{vacuum_energy_large_Nc}), derived from the large $N_c$ limit of a
bosonized Hamiltonian Eq. (\ref{bosonized}).

In the large $N_c$ limit, the fermion field $\psi$ can  be expressed as 
\begin{eqnarray}
\psi(x)=
\left(
\begin{array}{c}
\xi(x)\\ 
\eta^{\dagger}(x)
\end{array}
\right).
\label{second}
\end{eqnarray}
The 2-spinors $\xi$ and $\eta^{\dagger}$ are the annihilation operator of positive energy fermion 
and creation operator of negative energy fermion respectively. 
Let us define the state $\vert n_p, {\bar n}_p  \rangle$ in the momentum space by
\begin{eqnarray}
&& \xi_p \vert 0_p, {\bar n}_p  \rangle=0,
~~~ \xi^{\dagger}_p \vert 0_p, {\bar n}_p  \rangle=\vert 1_p, 
{\bar n}_p  \rangle, ~~~ \xi_p \vert 1_p, 
{\bar n}_p  \rangle=\vert 0_p, {\bar n}_p  \rangle,
~~~ \xi^{\dagger}_p \vert 1_p, {\bar n}_p  \rangle=0, 
\nonumber \\
&& \eta_p \vert n_p, 0_p  \rangle=0,
~~~ \eta^{\dagger}_p \vert n_p, 0_p  \rangle=\vert n_p, 1_p  \rangle, 
~~~ \eta_p \vert n_p, 1_p  \rangle=\vert n_p, 0_p  \rangle,
~~~ \eta^{\dagger}_p \vert n_p, 1_p  \rangle=0 .
\label{ninth}
\end{eqnarray}
The numbers $n_p$ and ${\bar n}_p$ take the values 0 
or 1 due to the Pauli principle. 
By definition, the up and down components of the fermion field
are decoupled in such a state $\vert n_p, {\bar n}_p  \rangle$. 
For the vacuum state of $H^{\mu}_{eff}$, we make an ansatz
\begin{eqnarray}
\vert \Omega_{eff} \rangle =  \sum_{n_p,{\bar n}_p, p} f_{n_p, {\bar n}_p} 
\vert n_p, {\bar n}_p  \rangle.
\label{tenth}
\end{eqnarray}

In the leading $N_c$ limit, the dominant contributions to the chiral condensate and quark number density are
\begin{eqnarray}
\langle {\bar \psi} \psi \rangle &=&
{\langle \Omega \vert \sum_x {\bar \psi}(x) \psi (x) \vert \Omega \rangle \over N_f N_s} 
\to  {\bar v}  
\nonumber\\
&=&
{1 \over N_f N_s} \sum_{n_p,{\bar n}_p, p} C_{n_p,{\bar n_p}} \langle n_{p},{\bar n_p}\vert {\bar \psi} \psi \vert n_p, {\bar n_p} \rangle 
= 2N_{c} (n+{\bar n} -1),
\nonumber\\
n_q &=& {\langle \Omega \vert \sum_x  \psi^{\dagger}(x) \psi (x) \vert \Omega \rangle \over 2N_c N_f N_s} -1 
\to {v^{\dagger} \over 2N_c}-1
\nonumber\\
&=&
{1 \over 2N_cN_f N_s} 
\sum_{n_p,{\bar n}_p, p} C_{n_p,{\bar n_p}} \langle n_{p},{\bar n_p}\vert \psi^{\dagger} \psi \vert n_p, {\bar n_p} \rangle -1
= n-{\bar n} .
\label{seventh1}
\end{eqnarray}
Here we denote $C_{n_p,{\bar n_p}}=f_{n_p, {\bar n}_p}^2$.
Using Eq. (\ref{vacuum_energy_large_Nc}) and Eq. (\ref{seventh1}), in the large $N_c$ limit 
we obtain the normalized vacuum energy
\begin{eqnarray}
\epsilon_{\Omega} &=& {E_{\Omega} \over 2N_cN_fN_s}
\nonumber\\
 &=& M_{chiral} \left(n+{\bar n} -1 \right) -\frac {Kdr^2}{a} \left(n- {\bar n} +1 \right)
\nonumber\\
 &+& \frac {Kdr^2}{a} \left(n^2 +{\bar n}^{2}+1-2{\bar n} \right) 
+\frac {Kd}{a}  \left(n+\bar {n} -2n {\bar n}-1 \right)
- \mu \left(n-\bar {n}+1 \right),
\label{energy_nn}
\end{eqnarray}
where the quark number $n$ and anti-quark number ${\bar n}$
\begin{eqnarray}
n &=& \langle n\rangle
= \sum_{n_p,{\bar n}_p, p} C_{n_p,{\bar n_p}} n_p,
\nonumber\\
{\bar n} &=& \langle {\bar n} \rangle
= \sum_{n_p,{\bar n}_p, p} C_{n_p,{\bar n}_p} {\bar n}_p
\end{eqnarray}
are constrained in the range of $[0,1]$ 
and determined by minimizing the vacuum energy.

For a generic nucleon operator $O_{Nucl}$ consisting of three quarks, the thermo mass is
\begin{eqnarray}
M_{Nucl} = \langle \Omega_{eff} \vert O_{Nucl}  H_{eff}^{\mu} O_{Nucl}^{\dagger} \vert \Omega_{eff} \rangle
-E_{\Omega}.
\label{Nucl}
\end{eqnarray} 
Under the mean-field approximation in the large $N_c$ limit, it becomes (see Appendix \ref{APPEN_NUC})
\begin{eqnarray}
M_{Nucl} = 3(A+B) - 3\mu .
\label{nucleon_mft}
\end{eqnarray}

\section{Phase structure at $T=0$ and $\mu \ne 0$}
\label{critical}

We now consider in the larger $N_c$ limit, the critical behavior of the system at $T=0$ and $\mu \ne 0$.
The ground state of the system corresponds to the lowest value of the vacuum energy.
At some given inputs of Wilson parameter $r$ and chemical potential $\mu$,
we can find the value of  $n$ and ${\bar n}$ when $\epsilon_{\Omega}$ Eq. (\ref{energy_nn}) is minimized.
The result is 
\begin{eqnarray}
n&=&\Theta(\mu-\mu_C),
\nonumber \\
{\bar n}&=&0.
\label{n_vacuum}
\end{eqnarray}
Here $\Theta(\mu-\mu_C)$ is the step function: it is 0 for $\mu<\mu_C$ and 1 for $\mu>\mu_C$,
where $\mu_C$ is the critical chemical potential
\begin{eqnarray}
\mu_C ={Kd \over a}  (1+2r^2).
\end{eqnarray}
Substituting Eq. (\ref{n_vacuum}) into Eq. (\ref{seventh1}), we obtain the chiral condensate and quark number density
\begin{eqnarray}
\langle {\bar \psi}\psi \rangle &=&2N_c \bigg(\Theta \left(\mu-\mu_C\right)-1\bigg),
\nonumber \\
n_q &=& \Theta(\mu-\mu_C).
\end{eqnarray}
There is clearly is a first order chiral phase transition. 
For $\mu<\mu_C$, the system is in the confinement phase with chiral-symmetry breaking.
For $\mu>\mu_C$, chiral symmetry is restored.

According to  Eqs. (\ref{consts}), (\ref{nucleon_mft}) and (\ref{n_vacuum}),
the thermo mass of the nucleon is
\begin{eqnarray}
M_{Nucl} = M_{Nucl}^{(0)} - 3\mu,
\label{thermo}
\end{eqnarray}
where 
\begin{eqnarray}
M_{Nucl}^{(0)}=3m_{dyn}^{(0)}
\label{nucleon_0}
\end{eqnarray}
is the nucleon mass at $\mu=0$, and
\begin{eqnarray}
m_{dyn}^{(0)}={Kd (1+r^2) \over a}
\end{eqnarray}
is the dynamical mass of quark at $\mu=0$.
From Eq. (\ref{thermo}), one sees that the nucleon thermo mass vanishes 
at $\mu=M_{Nucl}^{(0)}/3=m_{dyn}^{(0)}$, before the chiral
phase transition takes place. 
This is not surprising because Wilson fermions break explicitly the chiral symmetry.
The value of $\mu_C$ should coincide with  $M_{Nucl}^{(0)}/3$ when $r$ is very small, 
i.e., as in the case of naive or Kogut-Susskind fermions \cite{Gregory:1999pm}.

\section{Discussions}
\label{discussion}

In the preceding sections, we have investigated (d+1)-dimensional Hamiltonian lattice QCD 
at finite density with Wilson fermions in the strong coupling regime. 
We compute the vacuum energy, meson and nucleon masses, chiral condensate and quark number density. 
At finite chemical potential, there is an interplay between the bare fermion mass in the chiral limit and the chiral condensate,
which has to be determined self-consistently. 
The critical behavior of the system in the large $N_c$ limit is considered:
a first order chiral phase transition is found at $\mu=\mu_C$;
the nucleon thermo mass vanishes before the chiral transition takes place,
which is due to the explicit breakdown of chiral symmetry by Wilson fermions.

We have not yet specified the nature of the
chiral-symmetric phase for $\mu > \mu_C$.
Is it a QGP phase or a color-superconducting phase\cite{Rapp:1997zu,Alford:1997zt}?
Up to now, there has been no first principle investigation of such a phase in SU(3) gauge theory.
The answer to this question might be very important to
our understanding of the formation of the neutron star or quark star.

We also know that the strong coupling regime is far from the continuum limit.
One has to develop a new numerical method to study the continuum physics.
The Monte Carlo Hamiltonian method developed recently\cite{Jirari:1999jn,Huang:1999fn} might eventually be useful
for such a purpose.
We hope to discuss these interesting issues in the future.


\acknowledgments

We thank V. Azcoiti, S.H. Guo, H. Kr\"oger, V. Laliena, and M. Lombardo for useful discussions.
X.Q.L. is supported by the National Science Fund for Distinguished Young Scholars (19825117),
Key Project of National Science Foundation (10235040), 
Guangdong Natural Science Foundation (990212), 
National and Guangdong Ministries of Education, and
Hong Kong Foundation of the Zhongshan University 
Advanced Research Center.
Y.Z.F. is partly supported by the National Natural Science Foundation (10235040 and 10275098).

\appendix

\section{BOSONIZATION AND LINEARIZATION}
\label{APPEN_BOSON}

In Sect. \ref{results}, to bosonize the Hamiltonian Eq. (\ref{third}),
we introduce the operators $\Pi$, $\Pi^{\dagger}$, $V$ and $V^{\dagger}$, defined
by Eq. (\ref{twenty-seventh}). In Eq. (\ref{third}), there are terms having direct correspondence 
to these operators. In addition, there are also terms irrelevant to these operators.

In order that the operators in Eq. (\ref{twenty-seventh}) represent the appropriate mesons, 
they must satisfy the commutation relations for boson operators. 
However, a direct calculation shows
\begin{eqnarray}
\left[\Pi_{f_1,f_2} (x), \Pi_{f_2,f_1}^{\dagger}(x)\right] &=& {1 \over 2 {\bar v}} 
\left( {\bar \psi}_{f_1}(x)\psi_{f_1}(x)+{\bar \psi}_{f_2}(x)\psi_{f_2}(x) \right),
\nonumber\\
\left[V_{j,f_1,f_2}(x), V^{\dagger}_{l,f_2,f_1}(x)\right] &=& 
\delta_{jl} \frac{1}{2 \bar{v}}\left( {\bar \psi}_{f_1}(x)\psi_{f_1}(x)+{\bar \psi}_{f_2}(x)\psi_{f_2}(x) \right)
\nonumber\\
&&- \frac {1}{4\bar{v}}\left(\psi_{f_1}^{\dagger}(x)(\gamma_{j}\gamma_{l}-\gamma_{l}\gamma_{j})\psi_{f_1}(x)
+\psi_{f_2}^{\dagger}(x)(\gamma_{j}\gamma_{l}-\gamma_{l}\gamma_{j})\psi_{f_2}(x)\right),
\label{five}
\end{eqnarray}
which are not consistent with the commutation relations between the annihilation and creation operators for bosons.
Similar situation also appears in quantum theory of magnetization\cite{Mattis} and many-particle systems\cite{Fetter},
where a linearization prescription is used to simplify the theory. 
Using such a procedure, i.e.,  with the fermion bilinears 
in the r.h.s. of the commutation relations replaced by
the vacuum expectation value ${\bar \psi} (x)\psi (x) \to {\bar v}$ and for degenerate quarks, Eq. (\ref{five}) becomes
\begin{eqnarray}
\left[\Pi(x), \Pi^{\dagger}(x)\right] &\approx & \frac{1}{{\bar v}}
{\langle  \Omega_{eff} \vert \sum_x {\bar \psi} \psi \vert \Omega_{eff} \rangle \over N_fN_s} =1 ,
\nonumber\\
\left[V_{j}(x), V_{l}^{\dagger}(x)\right] &\approx & 
\delta_{jl}  
{\langle  \Omega_{eff} \vert \sum_x {\bar \psi} \psi \vert \Omega_{eff} \rangle \over N_fN_s{\bar v}}
-  
{\langle  \Omega_{eff} \vert \sum_x \psi^{\dagger}(x)(\gamma_{j}\gamma_{l}
-\gamma_{l}\gamma_{j})\psi(x) \vert \Omega_{eff} \rangle \over 2 N_fN_s {\bar v}}
=\delta_{jl},
\label{six}
\end{eqnarray}
which are the correct commutation relations for bosons.
As shown in Refs. \cite{Fang:2001ry,Guo:xa}, 
the linearization prescription leads to results consistent with Ref. \cite{Smit:1980nf},
where systematic $1/N_c$ expansion was used.

The four fermion terms in Eq. (\ref{third}), which have direct correspondence to the boson operators, are
\begin{eqnarray}
& &\sum_{x,k} \left( \psi_{f_1}^{\dagger}(x)\gamma_{5}\psi_{f_2} (x)\psi_{f_2}^{\dagger}(x+{\hat k})\gamma_{5}
\psi_{f_1} (x+{\hat k})
+\psi_{f_1}^{\dagger}(x)\gamma_{4}\gamma_{5}\psi_{f_2} (x)
 \psi_{f_2}^{\dagger}(x+{\hat k})\gamma_{4}\gamma_{5}\psi_{f_1} (x+{\hat k}) \right)
\nonumber \\
&=& -2{\bar v} \sum_{x,k} \left( \Pi_{f_1f_2}^{\dagger} (x) \Pi_{f_2f_1}^{\dagger}(x+{\hat k})
+\Pi_{f_1f_2}(x) \Pi_{f_2f_1} (x+{\hat k})\right),
\nonumber \\
& &\sum_{x,k} \left( \psi_{f_1}^{\dagger}(x)\gamma_{5}\psi_{f_2} (x)\psi_{f_2}^{\dagger}(x+{\hat k})\gamma_{5}
\psi_{f_1} (x+{\hat k})
-\psi_{f_1}^{\dagger}(x)\gamma_{4}\gamma_{5}\psi_{f_2} (x)
 \psi_{f_2}^{\dagger}(x+{\hat k})\gamma_{4}\gamma_{5}\psi_{f_1} (x+{\hat k}) \right)
\nonumber \\
&=& -2{\bar v} \sum_{x,k} \left( \Pi_{f_1f_2}^{\dagger} (x) \Pi_{f_2f_1}(x+{\hat k})
+\Pi_{f_1f_2}(x) \Pi_{f_2f_1}^{\dagger} (x+{\hat k})\right),
\nonumber \\
& &\sum_{x,k,j} \left( \psi_{f_1}^{\dagger}(x)\gamma_{4}\gamma_{j}\psi_{f_2} (x)
\psi_{f_2}^{\dagger}(x+{\hat k})\gamma_{4}\gamma_{j}\psi_{f_1} (x+{\hat k}) 
+ 
\psi_{f_1}^{\dagger}(x)\gamma_{j}\psi_{f_2} (x)
\psi_{f_2}^{\dagger}(x+{\hat k})\gamma_{j}\psi_{f_1} (x+{\hat k}) \right)
\nonumber \\
& & \times
\left(1-2\delta_{\vert k \vert, j}\right)
\nonumber \\
&=& 
-2{\bar v} \sum_{x,k,j}\left( V^{\dagger}_{jf_{1}f_{2}}(x)
V^{\dagger}_{jf_{2}f_{1}}(x+{\hat k})
+V_{jf_{1}f_{2}}(x)V_{jf_{2}f_{1}}(x+{\hat k})\right) \left(1-2\delta_{\vert k \vert, j}\right),
\nonumber \\
& &\sum_{x,k,j} \left( \psi_{f_1}^{\dagger}(x)\gamma_{4}\gamma_{j}\psi_{f_2} (x)
\psi_{f_2}^{\dagger}(x+{\hat k})\gamma_{4}\gamma_{j}\psi_{f_1} (x+{\hat k}) 
-
\psi_{f_1}^{\dagger}(x)\gamma_{j}\psi_{f_2} (x)
\psi_{f_2}^{\dagger}(x+{\hat k})\gamma_{j}\psi_{f_1} (x+{\hat k}) \right)
\nonumber \\
&=& 
2 {\bar v}
\sum_{x,k,j}
\left( V^{\dagger}_{jf_{1}f_{2}}(x)V_{jf_{2}f_{1}}(x+{\hat k})
+V_{jf_{1}f_{2}}(x)V^{\dagger}_{jf_{2}f_{1}}(x+{\hat k})\right).
\end{eqnarray}

For the bilinear $\sum_x {\bar \psi}(x)\psi(x)$ in Eq. (\ref{third}), 
a direct calculation shows
\begin{eqnarray}
\left[\frac {1}{2}\sum_{x'} {\bar \psi}(x')\psi(x'), \Pi_{f_1 f_2}(x)\right] 
&=& {1\over 4 \sqrt{-{\bar v}}} \bigg( 
\psi^{\dagger}_{f_{1}}(x) \gamma_4(1-\gamma_{4})\gamma_{5}\psi_{f_{2}}(x) 
\nonumber\\
&&-\psi^{\dagger}_{f_{1}}(x) \gamma_4(1-\gamma_{4})\gamma_{5} \gamma_4 \psi_{f_{2}}(x)
\bigg)
\nonumber\\
&=& -{1\over 2 \sqrt{-{\bar v}}} 
\psi^{\dagger}_{f_{1}}(x) (1-\gamma_{4})\gamma_{5}\psi_{f_{2}}(x)
=-\Pi_{f_1 f_2}(x).
\label{eight_1}
\end{eqnarray}
Similarly, one can also show
\begin{eqnarray}
\left[\frac {1}{2}\sum_{x'} {\bar \psi}(x')\psi(x'), \Pi_{f_2 f_1}^{\dagger}(x)\right] &=& \Pi_{f_2 f_1}^{\dagger}(x),
\nonumber\\
\left[\frac {1}{2}\sum_{x'} {\bar \psi}(x')\psi(x'), V_{jf_1 f_2}(x)\right] &=& -V_{jf_1 f_2}(x),
\nonumber\\
\left[\frac {1}{2}\sum_{x'} {\bar \psi}(x')\psi(x'), V^{\dagger}_{jf_2 f_1}(x)\right] &=& V^{\dagger}_{jf_2 f_1}(x) ,
\label{eight}
\end{eqnarray}

Eq. (\ref{eight_1}) and Eq. (\ref{eight}) imply that the bilinear $\sum_x {\bar \psi}(x)\psi(x)$ can be bosonized as
\begin{eqnarray}
\sum_{x}{\bar \psi}(x)\psi(x) \sim \sum_{x}\left( {\bar v} + 
2\Pi_{f_2 f_1}^{\dagger}(x)\Pi_{f_1 f_2}(x)+2\sum_j V^{\dagger}_{jf_2 f_1}(x)V_{jf_1 f_2}(x) \right).
\label{eleven}
\end{eqnarray}

For the four fermion term $\sum_{x}{\bar \psi}_{f_1}(x)\psi_{f_2}(x){\bar \psi}_{f_2}(x+{\hat k})\psi_{f_1}(x+{\hat k})$ in Eq. (\ref{third}),
we have
\begin{eqnarray}
& &\left[\frac {1}{2}\sum_{x'}{\bar \psi}_{f'_1}(x')\psi_{f'_2}(x'){\bar \psi}_{f'_2}(x'+{\hat k})\psi_{f'_1}(x'+{\hat k}), \Pi_{f_1f_2}(x)\right] 
\nonumber\\
&=& 
\sum_{x'} {\bar \psi}_{f'_1}(x')\psi_{f'_2}(x') \left[\frac {1}{2}{\bar \psi}_{f'_2}(x'+{\hat k})\psi_{f'_1}(x'+{\hat k}), \Pi_{f_1f_2}(x)\right]
\nonumber\\
&&+
\sum_{x'} \left[\frac {1}{2}{\bar \psi}_{f'_1}(x')\psi_{f'_2}(x'), \Pi_{f_1f_2}(x)\right] {\bar \psi}_{f'_2}(x'+{\hat k})\psi_{f'_1}(x'+{\hat k})
\nonumber\\
&\simeq&
- 2{\bar v} \Pi_{f_1f_2}(x) ,
\label{twelve_1}
\end{eqnarray}
where in the r.h.s., Eq. (\ref{eight_1}) and the linearization prescription have been used.
In addition, we also have
\begin{eqnarray}
\left[\frac {1}{2}\sum_{x'}{\bar \psi}_{f'_1}(x')\psi_{f'_2}(x'){\bar \psi}_{f'_2}(x'+{\hat k})\psi_{f'_1}(x'+{\hat k}), \Pi_{f_2f_1}^{\dagger}(x)\right] 
&\simeq& 2 {\bar v} \Pi_{f_2f_1}^{\dagger}(x) ,
\nonumber\\
\left[\frac {1}{2}\sum_{x'}{\bar \psi}_{f'_1}(x')\psi_{f'_2}(x'){\bar \psi}_{f'_2}(x'+{\hat k})\psi_{f'_1}(x'+{\hat k}), V_{jf_1f_2}(x)\right] 
&\simeq& - 2{\bar v} V_{jf_1f_2}(x),
\nonumber\\
\left[\frac {1}{2}\sum_{x'}{\bar \psi}_{f'_1}(x')\psi_{f'_2}(x'){\bar \psi}_{f'_2}(x'+{\hat k})\psi_{f'_1}(x'+{\hat k}), V^{\dagger}_{jf_2f_1}(x)\right] 
&\simeq& 2{\bar v} V_{jf_2f_1}^{\dagger}(x)  .
\label{nine}
\end{eqnarray}
Eq. (\ref{twelve_1}) and Eq. (\ref{nine}) also imply
that the four fermion term
$\sum_{x}{\bar \psi}_{f_1}(x)\psi_{f_2}(x){\bar \psi}_{f_2}(x+{\hat k})\psi_{f_1}(x+{\hat k})$ can be bosonized as
\begin{eqnarray}
&&\sum_{x}{\bar \psi}_{f_1}(x)\psi_{f_2}(x){\bar \psi}_{f_2}(x+{\hat k})\psi_{f_1}(x+{\hat k})
\nonumber\\
& \sim& \sum_x \left( {\bar v_2} 
 + 4 {\bar v} \Pi_{f_2f_1}^{\dagger}(x)\Pi_{f_1f_2}(x)+
4{\bar v} \sum_j V^{\dagger}_{jf_2f_1}(x)V_{jf_1f_2}(x) \right).
\label{twelve}
\end{eqnarray}

In the linear prescription, the four fermion operators in Eq. (\ref{third}), which are irrelevant 
to the operators in Eq. (\ref{twenty-seventh})
are replaced by their vacuum expectation values. For example,
\begin{eqnarray}
\psi^{\dagger} (x) \psi (x) &\sim & v^{\dagger},
\nonumber\\
\sum_{x,k} \psi^{\dagger}_{f_1} (x) \psi_{f_2} (x) \psi^{\dagger}_{f_2} (x+{\hat k}) \psi_{f_1} (x+{\hat k}) 
&\sim & 2dN_f N_s v_2^{\dagger},
\nonumber \\
\sum_{x,k,j}\left(r^2+\left(1-2\delta_{\vert k \vert,j} \right)\right){\bar \psi}_{f_1} (x)\sigma_j \psi_{f_2} (x) 
{\bar \psi}_{f_2}(x+{\hat k}) \sigma_j \psi_{f_1} (x+{\hat k}) 
&\sim & 2dN_fN_s {\bar v}_{2\sigma_j} ,
\nonumber \\
\sum_{x,k,j}\left(r^2-\left(1-2\delta_{\vert k \vert,j} \right)\right)\psi_{f_1}^{\dagger} (x) \sigma_j \psi_{f_2} (x) 
\psi_{f_2}^{\dagger}(x+{\hat k}) \sigma_j \psi_{f_1} (x+{\hat k}) 
& \sim & 2dN_fN_s v_{2\sigma_j}^{\dagger} .
%
\label{linear_v}
\end{eqnarray}

Collecting the above results,  we obtain Eqs. (\ref{bosonized}),
(\ref{energy_leadingNc}) and (\ref{thirty-first}).

\section{VACUUM ENERGY IN THE LARGE $N_c$ LIMIT}
\label{APPEN_VACUUM}

According to Eq. (\ref{v2}),  
\begin{eqnarray}
{\bar v}_2 & =&{1 \over 2dN_fN_s} 
\sum_{x,k,f_1,f_2} \bigg( \bigg\langle {\bar \psi}_{f_1,c_1} (x) \psi_{f_2,c_1} (x) \bigg\rangle_{eff}
\bigg \langle {\bar \psi}_{f_2,c_2}(x+{\hat k}) \psi_{f_1,c_2} (x+{\hat k}) \bigg \rangle_{eff}
\nonumber \\
&+& 
\bigg \langle \psi_{f_2} (x) {\bar \psi}_{f_2,c_2}(x+{\hat k}) \bigg \rangle_{eff}
\bigg \langle \psi_{f_1} (x+{\hat k}) {\bar \psi}_{f_1,c_1} (x) \bigg \rangle_{eff} \bigg) ,
\nonumber \\
v^{\dagger}_2 & =&{1 \over 2dN_fN_s} 
\sum_{x,k,f_1,f_2} \bigg( \bigg\langle \psi^{\dagger}_{f_1,c_1} (x) \psi_{f_2,c_1} (x) \bigg\rangle_{eff}
\bigg \langle \psi^{\dagger}_{f_2,c_2}(x+{\hat k}) \psi_{f_1,c_2} (x+{\hat k}) \bigg \rangle_{eff}
\nonumber \\
&+& 
\bigg \langle \psi_{f_2} (x) \psi^{\dagger}_{f_2,c_2}(x+{\hat k}) \bigg \rangle_{eff}
\bigg \langle \psi_{f_1} (x+{\hat k}) \psi^{\dagger}_{f_1,c_1} (x) \bigg \rangle_{eff} \bigg) ,
\label{vbar_largeNc}
\end{eqnarray}
where the color and flavor indexes for the fermion fields are explicitly specified
and summation over the repeated color index is implied. 
$\langle ... \rangle_{eff}$ stands for the expectation value taken in vacuum state $\vert \Omega_{eff} \rangle$.
Because 
\begin{eqnarray}
\bigg\langle {\bar \psi}_{f_1,c_1} (x) \psi_{f_2,c_1} (x) \bigg\rangle_{eff} & = &{\bar v} \delta_{f_1,f_2} \propto N_c ,
\nonumber \\
\bigg \langle \psi_{f_2,c_1} (x) {\bar \psi}_{f_2,c_2}(x+{\hat k}) \bigg \rangle_{eff} & \propto & \delta_{c_1,c_2} ,
\nonumber \\
\bigg\langle \psi^{\dagger}_{f_1,c_1} (x) \psi_{f_2,c_1} (x) \bigg\rangle_{eff} & = & v^{\dagger}\delta_{f_1,f_2} \propto N_c ,
\nonumber \\
\bigg \langle \psi_{f_2,c_1} (x) \psi^{\dagger}_{f_2,c_2}(x+{\hat k}) \bigg \rangle_{eff} & \propto & \delta_{c_1,c_2} ,
\label{Wick}
\end{eqnarray}
terms from Wick-contraction of bilinear at different lattice sites are subdominant for large $N_c$. Therefore, we have 
\begin{eqnarray}
{\bar v}_2 \longrightarrow {\bar v}^2 , ~~~~~~ v_2^{\dagger} \longrightarrow \left( v^{\dagger} \right)^2 ,
\label{v2_largeNc}
\end{eqnarray}
with ${\bar v}$ and $v^{\dagger}$ expectation value of fermion bilinears defined in Eq. (\ref{vv}).
Expectation values of other four fermion operators in Eq. (\ref{v2}) can be considered in the similar way,
\begin{eqnarray}
{\bar v}_{2\sigma_j} \propto \left({\rm Tr} \sigma_j \right)^2 = 0 , ~~~~~~ v_{2\sigma_j}^{\dagger} \propto \left({\rm Tr} \sigma_j \right)^2 = 0 
\label{v2j_largeNc}
\end{eqnarray}
in the large $N_c$ limit.

In the chiral limit Eq. (\ref{thirty-sixth}), 
$G_1$ defined in Eq. (\ref{thirty-forth}) scales as ${\bar v}K/N_c \propto K$
and $G_1/G_2$ doesn't depend on $N_c$.
The $N_f^2$ terms in Eq. (\ref{vacuum_energyN_c}) are subdominant because they are proportional to 
$K$, while
other non-vanishing terms of $E_{\Omega}^{(0)}$ in Eq. (\ref{vacuum_energyN_c}),
defined in Eq. (\ref{energy_leadingNc}) are dominant because they behave as 
${\bar v}K \propto KN_c$ or
$v^{\dagger} K \propto KN_c$. 
Using this fact and substituting Eqs. (\ref{v2_largeNc}) 
and (\ref{v2j_largeNc}) into Eq. (\ref{energy_leadingNc}),
we obtain Eq. (\ref{vacuum_energy_large_Nc}) from Eq. (\ref{vacuum_energyN_c}), i.e., the vacuum energy in the large $N_c$ limit.

\section{MEAN-FIELD APPROXIMATION IN THE LARGE $N_c$ LIMIT}
\label{APPEN_MEAN}

The mean-field approximation is a popular technique wildly used in quantum field theory with four fermion interactions 
(for example, to bilinearize the Nambu-Jona-Lasino model\cite{Nambu:tp,Buballa:1998ky})
and quantum theory of many-particle systems\cite{Fetter}.

In order to show how to derive Eq. (\ref{meanfield}) from Eq. (\ref{third}), 
let's look at the first and last of the four fermion terms 
in Eq. (\ref{third}) as an example.
Under the mean-field approximation, i.e., replacing a pair of fermion fields
by their vacuum expectation value, 
the first four fermion term in Eq. (\ref{third}) becomes
\begin{eqnarray}
 & & \sum_{x,k,f_1,f_2} \psi^{\dagger}_{f_1,c_1}(x)\psi_{f_2,c_1}(x)
\psi^{\dagger}_{f_2,c_2}(x+\hat{k})\psi_{f_1,c_2}(x+\hat{k})
\nonumber\\
 &\sim & \sum_{x,k,f_1,f_2}
\bigg(
\bigg\langle\psi^{\dagger}_{f_1,c_1}(x)\psi_{f_2,c_1}(x)\bigg\rangle_{eff}
\psi^{\dagger}_{f_2,c_2}(x+\hat{k})\psi_{f_1,c_2}(x+\hat{k})
\nonumber\\
& &+
\psi^{\dagger}_{f_1,c_1}(x)\psi_{f_2,c_1}(x)\bigg\langle
\psi^{\dagger}_{f_2,c_2}(x+\hat{k})\psi_{f_1,c_2}(x+\hat{k})\bigg\rangle_{eff}
\nonumber\\
& &+
\psi^{\dagger}_{f_1,c_1}(x) \bigg\langle \psi_{f_2,c_1}(x)
\psi^{\dagger}_{f_2,c_2}(x+\hat{k})\bigg\rangle_{eff} \psi_{f_1,c_2}(x+\hat{k})
\nonumber\\
& &+
\psi_{f_2,c_1}(x)\psi^{\dagger}_{f_2,c_2}(x+\hat{k}) \bigg\langle
\psi^{\dagger}_{f_1,c_1}(x)  \psi_{f_1,c_2}(x+\hat{k})\bigg \rangle_{eff} 
\bigg)  - 2dN_f N_sv_2^{\dagger}  .
\end{eqnarray}
Again, the color and flavor indexes for the fermion fields are explicitly specified
and summation over the repeated color index is implied. 
According to Eq. (\ref{Wick}), for large $N_c$,
Wick contractions of bilinear at different lattice sites give subdominant contribution,
therefore
\begin{eqnarray}
 & & \sum_{x,k,f_1,f_2} \psi^{\dagger}_{f_1,c_1}(x)\psi_{f_2,c_1}(x)
\psi^{\dagger}_{f_2,c_2}(x+\hat{k})\psi_{f_1,c_2}(x+\hat{k})
\nonumber\\
 &\approx & v^{\dagger} \sum_{x,k,f_1,f_2} \delta_{f_1,f_2} \bigg( 
\psi^{\dagger}_{f_2,c}(x+\hat{k})\psi_{f_1,c}(x+\hat{k})
+\psi^{\dagger}_{f_1,c}(x)\psi_{f_2,c}(x) \bigg) - 2dN_f N_s \left(v^{\dagger}\right)^2
\nonumber\\
 &=& v^{\dagger}\sum_{x,k,f}\left(\psi_{f,c}^{\dagger}(x)\psi_{f,c}(x)+\psi_{f,c}^{\dagger}(x+\hat{k})\psi_{f,c}(x+\hat{k})\right)
-2dN_f N_s \left(v^{\dagger}\right)^2
\nonumber\\
 &=& 4dv^{\dagger}\sum_{x} \psi^{\dagger}(x)\psi(x)  - 2dN_f N_s \left(v^{\dagger}\right)^2.
\label{term1}
\end{eqnarray}

Under the mean-field approximation in the large $N_c$ limit,
the dominant contribution to the last four fermion term in Eq. (\ref{third})  is
\begin{eqnarray}
& & \sum_{x,k,f_1,f_2} \psi^{\dagger}_{f_1,c_1}(x)\sigma_{j}\psi_{f_2,c_1}(x)
\psi^{\dagger}_{f_2,c_2}(x+\hat{k})\sigma_{j}\psi_{f_1,c_2}(x+\hat{k})
\nonumber\\
 &\sim & \sum_{x,k,f_1,f_2}
\bigg(\bigg\langle\psi^{\dagger}_{f_1,c_1}(x)\sigma_{j}\psi_{f_2,c_1}(x)
\bigg\rangle_{eff} \psi^{\dagger}_{f_2,c_2}(x+\hat{k})\sigma_{j}\psi_{f_1,c_2}(x+\hat{k})
\nonumber\\
& &+ \psi^{\dagger}_{f_1,c_1}(x)\sigma_{j}\psi_{f_2,c_1}(x)
\bigg\langle\psi^{\dagger}_{f_2,c_2}(x+\hat{k})\sigma_{j}\psi_{f_1,c_2}(x+\hat{k})\bigg\rangle_{eff}\bigg)  
- 2dN_f N_sv_{2 \sigma_j}^{\dagger}  
\nonumber\\
 &\propto & Tr(\sigma_{j}) = 0 .
\label{term2}
\end{eqnarray}
Treating other four fermion terms in the same way, 
one obtains  Eq. (\ref{meanfield}), i.e., the bilinear Hamiltonian $H_{MFA}$.

\section{NUCLEON MASS}
\label{APPEN_NUC}

Under the mean-field approximation in the large $N_c$ limit, Eq. (\ref{Nucl}) becomes
\begin{eqnarray}
M_{Nucl} = \langle \Omega_{eff} \vert O_{Nucl} \left( H_{MFA}-\mu N \right) O_{Nucl}^{\dagger}  \vert \Omega_{eff} \rangle
-E_{\Omega}.
\label{nucleon_mean}
\end{eqnarray} 
Substituting the bilinear Hamiltonian Eq. (\ref{meanfield}),  the fermion field Eq. (\ref{second}), 
and the particle number operator Eq. (\ref{seventh}) into Eq. (\ref{nucleon_mean}), we have
\begin{eqnarray}
M_{Nucl} &=& \langle\Omega_{eff}\vert O_{Nucl}
\left( A\sum_x {\bar \psi}(x) \psi(x)+(B-\mu)\sum_x\psi^{\dagger}(x)\psi(x)
+C \right) O_{Nucl}^{\dagger}\vert\Omega_{eff}\rangle-E_{\Omega}
\nonumber\\
 &=& \langle\Omega_{eff}\vert O_{Nucl}
\left( A\sum_x (\xi^{\dagger}\xi-\eta\eta^{\dagger})
+(B-\mu)\sum\limits_{x}(\xi^{\dagger}\xi+\eta\eta^{\dagger})
+C \right) O_{Nucl}^{\dagger}\vert\Omega_{eff}\rangle-E_{\Omega}
\nonumber\\
 &=& \langle\Omega_{eff}\vert O_{Nucl}
\left( A\sum\limits_{x} \xi^{\dagger}\xi+(B-\mu)\sum\limits_{x}\xi^{\dagger}\xi
\right) O_{Nucl}^{\dagger}\vert\Omega_{eff}\rangle,
\label{one}
\end{eqnarray}
where we have used the fact that the terms with $\eta^{\dagger}\eta$ and $C$ are canceled by $E_{\Omega}$.

Let us take the proton as an example. One can write the operator $O_{Nucl}^{\dagger}$ explicitly as
\begin{eqnarray}
O_{Nucl}^{\dagger}={1 \over \sqrt{18N_s}} \sum\limits_{x} \epsilon_{c_{1}c_{2}c_{3}}\xi_{c_1,u,1}^{\dagger}(x)
\left( \xi_{c_2,u,1}^{\dagger}(x)\xi_{c_3,d,2}^{\dagger}(x)
-\xi_{c_2,u,2}^{\dagger}(x)\xi_{c_3,d,1}^{\dagger}(x) \right),
\label{two}
\end{eqnarray}
where $c_{1},c_{2}$ and $c_{3}$ are the color indexes, $u$ and $d$ stand  respectively for the u-quark and d-quark, 
and $1$ and $2$ are the spin up and down indexes. Using the anti-commutation relationship for fermions
\begin{eqnarray}
\xi_{c,f,s}(x) \xi_{c',f',s'}^{\dagger}(x')
=\delta_{c,c'}\delta_{f,f'}\delta_{s,s'}\delta_{x,x'}- \xi_{c',f',s'}^{\dagger}(x')\xi_{c,f,s}(x),
\label{three}
\end{eqnarray}
one obtains
\begin{eqnarray}
\sum_x \xi^{\dagger}(x)\xi (x) O_{Nucl}^{\dagger}\vert\Omega_{eff}\rangle=3O_{Nucl}^{\dagger}\vert\Omega_{eff}\rangle.
\label{four}
\end{eqnarray}
Eq. (\ref{nucleon_mft}) is a consequence of Eq. (\ref{one}) and Eq. (\ref{four}).

At $\mu=0$, the nucleon mass equals to $3m_{dyn}^{(0)}$ (see Eq. (\ref{nucleon_0})), 
which agrees with Refs. \cite{Smit:1980nf,LeYaouanc:1985pq}.


\begin{thebibliography}{9}


\bibitem{Wilson:1974sk}
K. G. Wilson, Phys. Rev. D {\bf 10},  2445 (1974); 
Also, in New Phenomena in Sub-nuclear Physics, Erice Lectures 1975, 
A. Zichichi, ed. (Plenum, New York, 1977).


\bibitem{Hands:1999md}
S.~Hands, J.~B.~Kogut, M.~P.~Lombardo and S.~E.~Morrison,
Nucl.\ Phys.\ B {\bf 558},  327 (1999).

\bibitem{Aloisio:2000rb}
R.~Aloisio, V.~Azcoiti, G.~Di Carlo, A.~Galante and A.~F.~Grillo,
Nucl.\ Phys.\ B {\bf 606},   322 (2001).



\bibitem{Hasenfratz:1983ba}
P.~Hasenfratz and F.~Karsch,
Phys.\ Lett.\ B {\bf 125},  308 (1983).


\bibitem{Palumbo:2002jf}
F.~Palumbo,
Nucl.\ Phys.\ B {\bf 645}, 309 (2002).



\bibitem{Fodor:2002hs}
Z.~Fodor and S.~D.~Katz,
Phys.\ Lett.\ B {\bf 534},  87 (2002).
J. High Energy Phys. {\bf 0203}, 014 (2002).
Heavy Ion Phys.\  {\bf 18}, 41 (2003);
Phys.\ Lett.\ B {\bf 568}, 73 (2003).



\bibitem{deForcrand:2002ci}
P.~de Forcrand and O.~Philipsen,
Nucl.\ Phys.\ B {\bf 642},  290 (2002).


\bibitem{D'Elia:2002pj}
M.~D'Elia and M.~P.~Lombardo,
arXiv:hep-lat/0205022.


\bibitem{Gregory:1999pm}
E.~B.~Gregory, S.~H.~Guo, H.~Kr\"oger and X.~Q.~Luo,
Phys.\ Rev.\ D {\bf 62},  054508 (2000).


\bibitem{Luo:2000xi}
X.~Q.~Luo, E.~B.~Gregory, S.~H.~Guo and H.~Kr\"oger,
arXiv:hep-ph/0011120.



\bibitem{LeYaouanc:1987ff}
A.~Le Yaouanc, L.~Oliver, O.~Pene, J.~C.~Raynal, M.~Jarfi and O.~Lazrak,
Phys.\ Rev.\ D {\bf 37},  3691 (1988).


\bibitem{Aloisio:1999nk}
R.~Aloisio, V.~Azcoiti, G.~Di Carlo, A.~Galante and A.~F.~Grillo,
Nucl.\ Phys.\ B {\bf 564},  489 (2000).

\bibitem{Rothe}    H. Rothe, ``Lattice Gauge Theories - an Introduction", World Scientific Press (1999), and references therein.


\bibitem{Montvay}  I. Montvay, G. Munster, ``Quantum Fields on a Lattice", Cambridge University Press (1994), and references therein.

\bibitem{Luo:fc}
X.~Q.~Luo,
Z.\ Phys.\ C {\bf 48},  283 (1990).

\bibitem{Chen:ej}
Q.~Z.~Chen and X.~Q.~Luo,
Phys.\ Rev.\ D {\bf 42},  1293 (1990).



\bibitem{Luo:1990dm}
X.~Q.~Luo,
Commun.\ Theor.\ Phys.\  {\bf 16},  505 (1991).


\bibitem{Azcoiti:1995bx}
V.~Azcoiti, G.~DiCarlo, A.~Galante, A.~Grillo and V.~Laliena,
Phys.\ Rev.\ D{\bf 53},  5069 (1996).


\bibitem{Smit:1980nf}
J.~Smit,
Nucl.\ Phys.\ B {\bf 175},   307 (1980).


\bibitem{Fang:2001ry}
Y.~Z.~Fang and X.~Q.~Luo,
Int.\ J.\ Mod.\ Phys.\ A {\bf 16},  4499 (2001).


\bibitem{Umino:2001ie}
Y.~Umino,
Phys.\ Rev.\ D {\bf 66},  074501 (2002).


\bibitem{Luo:kk}
X.~Q.~Luo and Q.~Z.~Chen,
Phys.\ Rev.\ D {\bf 46}, 814 (1992).
 
\bibitem{Guo:xa}
S.~H.~Guo, Q.~H.~Chen, J.~M.~Liu and L.~Hu,
Commun.\ Theor.\ Phys.\  {\bf 3}, 575 (1984).






\bibitem{Rapp:1997zu}
R.~Rapp, T.~Schafer, E.~V.~Shuryak and M.~Velkovsky,
Phys.\ Rev.\ Lett.\  {\bf 81},   53 (1998).


\bibitem{Alford:1997zt}
M.~G.~Alford, K.~Rajagopal and F.~Wilczek,
Phys.\ Lett.\ B {\bf 422},   247 (1998).





\bibitem{Jirari:1999jn}
H.~Jirari, H.~Kroger, X.~Q.~Luo and K.~J.~Moriarty,
Phys.\ Lett.\ A {\bf 258},  6 (1999).


\bibitem{Huang:1999fn}
C.~Q.~Huang, X.~Q.~Luo, H.~Kroger and K.~J.~Moriarty,
Phys.\ Lett.\ A {\bf 299},   483 (2002).



\bibitem{Mattis}
D.C. Mattis, The Theory of Magnetism, Harper and Row, New York (1965).

\bibitem{Fetter}
A. Fetter and J. Walecka, Quantum Theory of Many-Particle Systems, McGraw-Hill, Inc. (1971).


\bibitem{Nambu:tp}
Y.~Nambu and G.~Jona-Lasinio,
Phys.\ Rev.\  {\bf 122},  345 (1961).


\bibitem{Buballa:1998ky}
M.~Buballa and M.~Oertel,
Nucl.\ Phys.\ A {\bf 642}, 39 (1998).

\bibitem{LeYaouanc:1985pq}
A.~Le Yaouanc, L.~Oliver, O.~Pene and J.~C.~Raynal,
Phys.\ Rev.\ D {\bf 33}, 3098 (1986).


\end{thebibliography}
\end{document}